\documentclass[runningheads]{llncs}
\usepackage{times}
\usepackage{latexsym}
\usepackage{graphicx}
\usepackage{url}

\usepackage{xcolor}
\usepackage{multirow}
\usepackage{colortbl}
\usepackage{booktabs}
\usepackage{amsmath}
\usepackage{bm}
\usepackage{amsfonts}

\DeclareMathOperator*{\simf}{ATS}
\DeclareMathOperator*{\msimf}{MATS}
\DeclareMathOperator*{\tff}{TF}

\DeclareMathOperator*{\tlf}{TL}

\DeclareMathOperator*{\bertcat}{BERT_\text{CAT}}
\DeclareMathOperator*{\bertdot}{BERT_\text{DOT}}

\title{Mitigating the Position Bias of Transformer Models in~Passage~Re-Ranking}

\author{Sebastian Hofst{\"a}tter\inst{1} \and Aldo Lipani\inst{2} \and Sophia Althammer\inst{1} \and Markus Zlabinger\inst{1} \and Allan Hanbury\inst{1}}

\authorrunning{S. Hofst{\"a}tter et al.}
\institute{TU Wien, Vienna, Austria \\ \email{\{\textit{first.last}\}@tuwien.ac.at} \and University College London, United Kingdom,\\ \email{aldo.lipani@ucl.ac.uk}}

\date{}

\begin{document}
\maketitle

\begin{abstract}
Supervised machine learning models and their evaluation strongly depends on the quality of the underlying dataset. When we search for a relevant piece of information it may appear anywhere in a given passage. However, we observe a bias in the position of the correct answer in the text in two popular Question Answering datasets used for passage re-ranking. The excessive favoring of earlier positions inside passages is an unwanted artefact. This leads to three common Transformer-based re-ranking models to ignore relevant parts in unseen passages. More concerningly, as the evaluation set is taken from the same biased distribution, the models overfitting to that bias overestimate their true effectiveness. In this work we analyze position bias on datasets, the contextualized representations, and their effect on retrieval results. We propose a debiasing method for retrieval datasets. Our results show that a model trained on a position-biased dataset exhibits a significant decrease in re-ranking effectiveness when evaluated on a debiased dataset. We demonstrate that by mitigating the position bias, Transformer-based re-ranking models are equally effective on a biased and debiased dataset, as well as more effective in a transfer-learning setting between two differently biased datasets.
\end{abstract}

\section{Introduction}

Datasets used to train neural network models are subject to a range of biases, which might constitute unwanted artefacts that should not be incorporated in the trained model \cite{gururangan-etal-2018-annotation}. Multiple studies showed that in the ad-hoc retrieval of full documents the text location is of relevant importance, such as the beginning in news articles \cite{catena2019,wu2019investigating} or general web search \cite{hofstatter2020fine}. In contrast, in this study we specifically probe positional bias in passage collections that are not linked to the previously studied full document relevance distributions. We operate on the assumption, based on the findings of the annotation study of TREC'19 Deep Learning data \cite{trec2019overview} by Hofst{\"a}tter et al. \cite{hofstatter2020fine}, that inside a passage (made up of a few sentences) no word position is supposed to be explicitly favored when matching query and passage sequences. 

Transformer-based neural re-ranking models, especially models based on the large-scale pre-trained BERT model \cite{devlin2018bert}, have shown a significant improvement in ad-hoc retrieval, where a natural language question is asked by the user and a set of passages is retrieved \cite{nogueira2019passage,macavaney2019}. In this study we evaluate three state-of-the-art Transformer ranking models with varying characteristics: \textbf{ 1) BERT$_\textbf{CAT}$} \cite{nogueira2019passage} using BERT with query and passage concatenation, \textbf{2) BERT$_\textbf{DOT}$} \cite{xiong2020approximate} using a dot-product between query and passage BERT classification (CLS) vectors and \textbf{3) TK} \cite{Hofstaetter2020_ecai}, a lightweight Transformer-Kernel model that does not require pre-training. Each of the three architectures exhibits different strengths and weaknesses, which we describe in Section \ref{sec:background}.

In the Transformer-architecture, positional information is induced through absolute position information provided by a positional encoding \cite{vaswani2017attention}. This positional encoding is added to each non-contextualized representation in a sequence before applying the self-attention. If a bias favoring certain positions in a text exists the Transformer may implicitly incorporate this bias in its word representation as Transformers tend to learn positional information \cite{yang-etal-2019-assessing}. To our knowledge, the connection between the explicit positional information of the Transformer and positional artefacts in common retrieval collections has not been studied before.

Traditional IR datasets contain relevance judgements for query-document pairs, where a single judgement covers the full document. In contrast to that, 
QA datasets contain exact location spans of the answer or an answer text that can be partly matched to a position in the document. In our work, we utilize two widely used QA datasets: MS MARCO \cite{msmarco16} and SQuAD 2.0 \cite{rajpurkar2016squad}.
Both datasets are converted to retrieval collections, by setting paragraphs that were selected to contain the answer as a relevant paragraph for a question. We observe that for the MS MARCO dataset the positions of the mapped answers strongly favor earlier positions in the paragraphs, while the SQuAD 2.0 dataset is more balanced although not completely bias free. The evaluation set is taken from the same distribution, therefore the evaluation is also biased and models overfitting to that bias overestimate their true effectiveness. In the case of MS MARCO this bias is especially concerning as it -- because of its size -- became the defacto standard collection in the neural re-ranking community, including as base retrieval training for transfer learning \cite{li2020parade,yilmaz2019cross}.

We propose to create unbiased versions of the datasets by switching the first and second parts of a passage around a randomly selected position. This approach does not affect the relevance judgements, since they are on a passage level, and allows us to train unbiased re-ranking models as well as to measure the true effectiveness of re-ranking approaches, since relevant matches might now occur in every part of the passage.

We analyze passage term representations to study the position bias induced in Transformer based contextualization and answer \textit{\textbf{RQ1} How can we measure the degree of position bias in the passage representations?} We propose a new metric to measure the mean average term similarity (MATS) per positional delta of all terms in the collection to investigate whether the term representations are independent of the positional encoding or not. 

To understand the effects of our debias augmentation in conjunction with Transformer models we further study the following questions:

\noindent
\textit{\textbf{RQ2} What effect has the debiasing on the evaluation of Transformers?}

We evaluate the effectiveness of our modifications on the original, as well as the debiased collections. We find that all three models perform better on the original (biased) evaluation, but their effectiveness drops substantially on a debiased evaluation set. 
\newpage
\noindent
\textit{\textbf{RQ3} Does a debiased training result in better generalization?}

Training on an unbiased collection shows much more robust results across the evaluated collections and models, which we view as a a more accurate indicator for their actual effectiveness.

\noindent
\textit{\textbf{RQ4} Do we observe differences in transfer-learning, based on debiased pre-training?}

We demonstrate the usefulness of mitigating bias in the learned representations in the scenario of transfer learning between differently biased collections. We use the larger MS MARCO to pre-train our model variants, before fine-tuning the models on  SQuAD 2.0. The bias-mitigated pre-training shows more effective results in the fine-tuning, than starting with a biased pre-training.

The contributions of this work are as follows:
\begin{itemize}
   \item We measure the positional bias of judgments in two popular Open-QA passage retrieval collections and propose a method to debias the collections;
    \item We show how three different Transformer-based re-ranking models learn to incorporate the position bias;
    \item We demonstrate the importance of mitigating the position bias with debiased evaluation sets and the benefit of debiasing in transfer learning between collections.
    \item We publish the source code of our work at: \\ \textit{\scriptsize\url{github.com/sebastian-hofstaetter/transformer-kernel-ranking}}
\end{itemize}

\section{Background}
\label{sec:background}
In this section we first describe the Transformer architecture, followed by the three Transformer-based passage re-ranking models we employ in this study.
\vspace{-0.3cm}
\subsection{Transformer}
\label{sec:transformer}
\vspace{-0.1cm}
The Transformer-layer \cite{vaswani2017attention} is a versatile building block for different architectures. In our work we use an encoder structure to encode a sequence and output contextualized representations of this sequence.
The Transformer architecture incorporates a natural algorithmic bias on the position of a term in a sequence, because it adds a positional encoding to its input sequence. Vaswani et~al.~\cite{vaswani2017attention} use overlapping sinusoidal-waves per dimension, forming an equidistant relationship among neighbouring terms, whereas Devlin et al. \cite{devlin2018bert} employ a trainable positional embedding for BERT.
This positional encoding is important since the Transformer otherwise would be entirely invariant to sequence ordering.
However, adding the positional encoding directly to the input means that absolute positional information is retained in the output sequence. Each encoding is unique to a position of the input sequence. Based on the provided training examples, the Transformer may tend to learn position-biased representations. 

In this paper we define the Transformer as the sequential use of $n$ Transformer-layers ($\tlf$s) as:
\begin{equation}
\begin{aligned}
    s_{1:m}^{(1)} & = \tlf(s_{1:m})\\
    s_{1:m}^{(n)} & = \tlf(s_{1:m}^{(n-1)})\\
    \tff(s_{1:m} + e_{1:m}) & = s_{1:m}^{(n)}
\end{aligned}
\end{equation}
where $s_{1:m}$ is the sequence of input embeddings, $e_{1:m}$ is the positional encoding.
We call this sequence of recursive applications $\tff$.

\subsection{BERT$_\textbf{CAT}$ Ranking Model}
\label{sec:bert_cat_model}

First proposed by Nogueira et al. \cite{nogueira2019passage} the BERT$_\text{CAT}$ approach has become a common way of utilizing the BERT pre-trained Transformer model in a re-ranking scenario \cite{macavaney2019,yilmaz2019cross}. It uses the capability of the BERT pre-training approach to compute the relationship of two concatenated sequences, separated by a special SEP token and depending on the BERT version a sequence embedding. The BERT architecture is a simple Transformer model (TF), the effectiveness comes from the masked language and next sentence prediction pre-training. In the BERT$_\text{CAT}$ ranking model the query (${q}_{1:m}$) and passage (${p}_{1:n}$) sequences as well as BERT's special tokens are concatenated (where $;$ is the concatenation operator) and after the TF computation we select only the first vector of the output sequence (which has been initialized with the special CLS token) and score this pooled representation with a single linear layer ($W_s$):
\begin{equation}
\begin{aligned}
    \bertcat({q}_{1:m},{p}_{1:n}) & = \tff([\text{CLS};{q}_{1:m};\text{SEP};{p}_{1:n}])_1 * W_s
\end{aligned}
\end{equation}

BERT$_\text{CAT}$ is the current state-of-the art in terms of effectiveness, however it requires substantial compute at query time and increases the query latency by seconds \cite{Hofstaetter2019_osirrc}. Therefore, we also feature additional models that provide a more balanced efficiency-effectiveness tradeoff.

\subsection{BERT$_\textbf{DOT}$ Ranking Model}
\label{sec:bert_dot_model}

In contrast to the full-interaction BERT$_\text{CAT}$ model, that requires a full online computation of all selected passages, the BERT$_\text{DOT}$ model only matches a single CLS vector of the query with a single CLS vector of a passage \cite{xiong2020approximate,luan2020sparse}. This makes it possible to pre-compute contextualized representations for all passages in our index, as well as to employ a vector-based nearest neighbour retrieval approach.

The BERT$_\text{DOT}$ model, with $\cdot$ as the dot product operator, is formalized by two independent $\tff$ computations (and their pooled representations by selecting the first vector output) as follows:
\begin{equation}
\begin{aligned}
    \bertdot({q}_{1:m},{p}_{1:n}) & = \tff([\text{CLS};{q}_{1:m}])_1 \cdot \tff([\text{CLS};{p}_{1:n}])_1
\end{aligned}
\end{equation}

BERT$_\text{DOT}$ brings strong query time improvements (a few milliseconds latency per query) over BERT$_\text{CAT}$, however it still requires the full BERT pre-computation of all indexed passages, which can be very costly depending on the collection size. 

\subsection{TK Ranking Model}
\label{sec:tk_base}

The TK model \cite{Hofstaetter2020_ecai}, while also utilizing Transformers, is not based on BERT pre-training, rather it uses shallow Transformers atop word embeddings followed by an explicit term-by-term interaction matrix and scoring with kernel-pooling \cite{Xiong2017}. In contrast to the BERT approaches TK offers us great control to probe the individual term representations, as it splits the representation learning and their interactions in two distinct parts. 

The first part of TK is learning contextualized representations. 
TK independently contextualizes query (${q}_{1:m}$) and passage (${p}_{1:n}$) sequences based on pre-trained word embeddings, where the intensity of the contextualization (with $\tff$) is regulated by a gate ($\alpha$):
\begin{equation}
\begin{aligned} 
\hat{q}_i &= q_i * \alpha + \tff(q_{1:m})_i * (1 - \alpha) \\
\hat{p}_i &= p_i * \alpha + \tff(p_{1:n})_i * (1 - \alpha)
\label{eq:representation}
\end{aligned}
\end{equation}

The two resulting sequences $\hat{q}_{1:m}$ and $\hat{p}_{1:n}$ interact in a  match-matrix with a cosine similarity per term pair and each similarity is then activated by a set of RBF-kernels \cite{Xiong2017}:
\begin{equation}
K^{k}_{i,j} = \exp \left(-\frac{\left(\cos(\hat{q_i},\hat{p_j})-\mu_{k}\right)^{2}}{2 \sigma^{2}}\right)
\end{equation}
Kernel-pooling is conceptually a soft-histogram, which counts the number of occurrences of certain similarities. Each kernel focuses on a fixed similarity range with center $\mu_{k}$ and width of $\sigma$. Each kernel results in a matrix $K \in \mathbb{R}^{|q| \times |p|}$.

These kernel activations are then summed, first by the passage term dimension $j$, log-activated, and then the query dimension is summed resulting in a single score per kernel. The final score is calculated by a weighted sum using the linear layer $W_s$:
\begin{equation}\label{eq:rsv}
s = \bigg(\sum_{i=1}^{|q|} \log\left( \sum_{j=1}^{|p|} K^{k}_{i,j} \right) \bigg) W_s
\end{equation}
The kernel-pooling technique is position-independent, as every activation for position $j$ is summed without a weighting them, which allows us to isolate the positional analysis in the Transformer in Section \ref{sec:transformer_analysis}.

\section{Experiment Design}
\label{sec:experiments}
For the first stage indexing and retrieval we use the Anserini toolkit \cite{Yang2017} to compute the initial ranking lists with BM25, which we use to generate training and evaluation inputs for the neural models. For our neural re-ranking training and inference we use PyTorch~\cite{pytorch2017} and AllenNLP~\cite{Gardner2017AllenNLP}. We tokenize the text with the fast BlingFire library\footnote{\textit{github.com/microsoft/BlingFire}}. 
As proposed for the MS MARCO dataset \cite{msmarco16} we evaluate our neural re-ranking systems using mean reciprocal rank (MRR), normalized discounted cumulative gain (nDCG), and recall (Recall).


For the BERT-based models we use the 6-layer DistilBERT \cite{sanh2019distilbert} pre-trained weights and the Adam \cite{kingma2014adam} optimizer with a learning rate of $7*10^{-6}$. For TK we use pre-trained GloVe~\cite{pennington2014glove} word embeddings with 300 dimensions\footnote{42B CommonCrawl: \textit{nlp.stanford.edu/projects/glove/}} and Adam with a learning rate of $10^{-4}$ for word embeddings and contextualization layers, $10^{-3}$ for the kernel-pooling weights.

For the Transformer layers in TK we evaluate 2 layers each with 16 attention heads with size 32 and a feed-forward dimension of 100. For kernel-pooling we set the number of kernels to $11$ with the mean values of the Gaussian kernels varying from $-1$ to $+1$, and standard deviation of $0.1$ for all kernels. We use the same sinusoidal positional encodings as Vaswani et al. \cite{vaswani2017attention}, for the document encodings we shift the start position by 500 to distinguish them from the query encodings. 

We train all neural models with a pairwise hinge loss and a batch size of 32. The re-ranking depth for each model instance is tuned on the best mean nDCG@10 of the validation set, as part of an early stopping strategy. For MS MARCO we evaluate a re-ranking depth until 1000 and for SQuAD up to 100.

\section{Dataset Analysis \& Debiasing}
\label{sec:dataset_analysis}

\begin{table}[t!]
    \centering
    \caption{Collection statistics}
    \label{tab:collection_stats}
    {
    \setlength\tabcolsep{6pt}
    \begin{tabular}{lr!{\color{lightgray}\vrule}rrr}
       \toprule
       \multirow{2}{*}{\textbf{Collection}} & 
       \multirow{2}{*}{\textbf{\# Docs.}} &
       \multicolumn{3}{c}{\textbf{\# Queries}} \\
       & &\textbf{Train} & \textbf{Val.} & \textbf{Test} \\ \midrule
       \textbf{MS MARCO}  & 8,841,823  & 502,939 & 6,980 & 48,598  \\
       \textbf{SQuAD 2.0}     & 20,239 &  86,821 & 5,000  &  5,928  \\
        \bottomrule
    \end{tabular}}
\end{table}

To better understand the neural models, we first need to look at the source of the position bias of the training and evaluation data, specifically the distribution of answer positions in our QA-datasets.

\subsection{Dataset Analysis}

The question answering task is strongly linked to ad-hoc information retrieval, as IR provides the first stage of selecting potential candidate passages that contain the natural language answer, that should be presented to a user. In addition to traditional relevance judgements, that cover full documents, the QA datasets also contain short answer strings or exact spans pointing to the answer in a passage.

Using QA datasets to evaluate the retrieval portion of the QA pipeline offers us the unique opportunity of inspecting the answer position, which gives us an insight into the positional importance inside the relevant passages. For SQuAD 2.0 we follow the approach done for MS MARCO \cite{msmarco16} and set a passage as relevant to a query if the passage is connected to the answer. We provide an overview of the size of our collections in Table \ref{tab:collection_stats}, where we observe that MS MARCO is a much larger collection than SQuAD.

In Figure \ref{fig:collection_plots} we show the distribution of the QA-answer start positions in their respective relevant passages for the training sets of MS MARCO and SQuAD. To determine the answer positions, we matched the available answer tokens to the passage tokens of the selected passages for both collections and counted all matches. For MS MARCO we omitted answers that could not directly be matched in the passage. In this figure, it is evident that the answer positions in the MS MARCO dataset strongly favor earlier positions in the paragraphs. MS MARCO was created in a retrieval setting, where annotators were given a question and a list of 10 possible paragraphs to judge, which may have favoured passages with answers appearing early in the text. On the other hand SQuAD 2.0, for which annotators were asked to create questions based on a given passage, is relatively unbiased, as the distribution of answer spans in the paragraphs is more uniform.

\subsection{Debiasing the Passage Datasets}

\begin{figure*}[t!]
    \centering
    \includegraphics[clip,trim={0cm 0cm 0cm 0cm} ,width=0.7\textwidth]{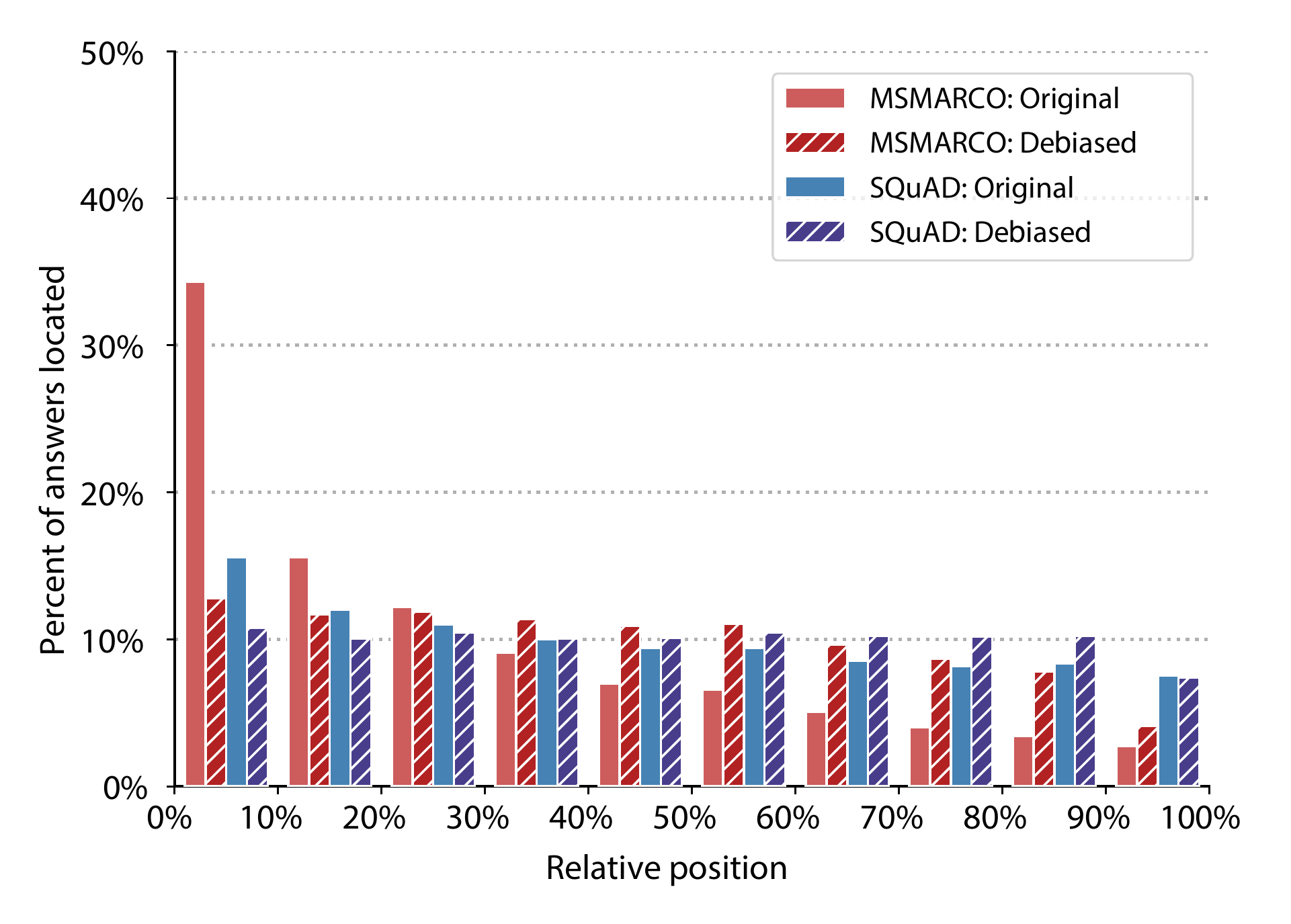}
    \vspace{-0.4cm}
    \centering
    \caption{QA collection in-passage relative answer positions}
    \label{fig:collection_plots}
    \vspace{-0.2cm}
\end{figure*}

We have established that MS MARCO answers excessively favor the beginning of a passage, while SQuAD does not. To explicitly study this phenomenon, Hofst{\"a}tter et al. \cite{hofstatter2020fine} conducted a fine-grained relevance position study. They found, that if annotators are shown only one query passage pair at a time, annotators select answers uniformly across passages. As we simply cannot re-annotate a collection of the size of MSMARCO with hundreds of thousands of queries, we apply an automatic debiasing method to the existing collections.

For each passage ${p}_{1:n}$ in the collection we create a \emph{debiased} instance $\tilde{p}_{1:n}$, for which we generate a random number $r \in \{1, \dots, n\}$, slice the word sequence at the $r^{th}$ index, switch and concatenate the two sub-sequences again:
\begin{equation}
\tilde{p}_{1:n} = \left[p_{r:n}\,;\,p_{1:r-1} \right]
\end{equation}
As shown in Figure \ref{fig:collection_plots} this approach produces near uniformly distributed relative answer positions for both collections. This approach is minimally invasive as it only breaks the contextualization at a single point per passage, without the need for additional annotations. In a pilot study we also experimented with sentence splitting based rotation, however we found that in the the MSMARCO web-page collection too many passages do not contain punctuation and therefore the sentence split approach does not produce uniform answer positions.

\section{Transformer Bias Analysis}
\label{sec:transformer_analysis}

\begin{table}[t!]
    \centering
    \caption{$\msimf$ statistics for TK's contextualized passage vectors. \textit{Lower MATS means less position bias.}}
    \label{tab:position_bias_stats}
    \vspace{-0.2cm}
    \setlength\tabcolsep{6pt}
    {
    \begin{tabular}{l!{\color{lightgray}\vrule}rr!{\color{lightgray}\vrule}rr}
       \toprule
       \multicolumn{1}{c!{\color{gray}\vrule}}{\multirow{2}{*}{\textbf{Training}}}&
       \multicolumn{2}{c!{\color{lightgray}\vrule}}{\textbf{MS MARCO}}&
       \multicolumn{2}{c}{\textbf{SQuAD}} \\
       &  MATS &  Std.dev. & MATS &  Std.dev.  \\
        \midrule
        {Original} &   0.176  & 0.046 & 0.056  & 0.014\\
        {Debiased}  &   \textbf{0.021}  & \textbf{0.006} & \textbf{0.007}  & \textbf{0.002} \\
        \bottomrule
    \end{tabular}}
\end{table}

In this section we probe term-wise Transformer representations to determine their bias across positions. Both BERT model variants incorporate their scoring decision mechanism inside the Transformer layers and only use the CLS vector representation, hiding individual term interactions inside the model. The TK model on the other hand utilizes every passage term representation in the cosine match matrix, which allows us to decouple the Transformer layers from the relevance scoring and analyze the passage term representations of a trained model on their own.  

We now discuss \textit{\textbf{RQ1} How can we measure the degree of position bias in the passage representations?} by analyzing the implicit bias of the absolute position of a term in a sequence. If a contextualized vector contains enough information about the original position, then a bias is measurable when we compare different vectors of the same term. We propose to compare the cosine similarity of the contextualized representations $r$ between occurrences of the same term $t$ across different passages computing the average term similarity ($\simf$) at distance $\Delta a$ for all terms  in the collection $t \in \mathcal{T}$. This is formalized as follows:

\begin{equation}
\begin{aligned}
\simf(\Delta a) &= \frac{1}{|\mathcal{T}|} \sum_{t \in \mathcal{T}} \frac{1}{|C_{t,\Delta a}|} \sum_{(r^{t}_{a_1},r^{t}_{a_2}) \in C_{t,\Delta a}}     \cos(r^{t}_{a_1},r^{t}_{a_2}) \\
C_{t,\Delta a} & = \big\{ (r^{t}_{a_1},r^{t}_{a_2}) \big| \Delta a = |a_1 - a_2| , (t_{a_1} , t_{a_2}) \in \mathcal{C}  \big\}
\end{aligned}
\end{equation}

where $r^{t}_{a_1}$ is the representation of term $t$ at absolute position $a_1$. The set $C_{t,\Delta a}$ is a set of all couples of representations of term $t$, which occur in the passages with a distance between their absolute positions of $\Delta a = |a_1 - a_2|$ in the collection $\mathcal{C}$.
The mean $\simf$ difference to the first point ($\msimf$) is computed as:
\begin{equation}
\begin{aligned}
\msimf = \frac{1}{\max(\Delta a) - 1} \sum_{i=1}^{\max(\Delta a)} \simf(0) - \simf(i) 
\end{aligned}
\end{equation}

$\msimf$ aggregates $\simf$ across all available positions in the passages and allows us to formally compare the different distributions. In Table \ref{tab:position_bias_stats} we show TK's $\msimf$ for both collections.

\begin{figure*}[t]
   \includegraphics[trim={0.7cm 0.7cm 0.2cm 0.4cm},width=1\textwidth]{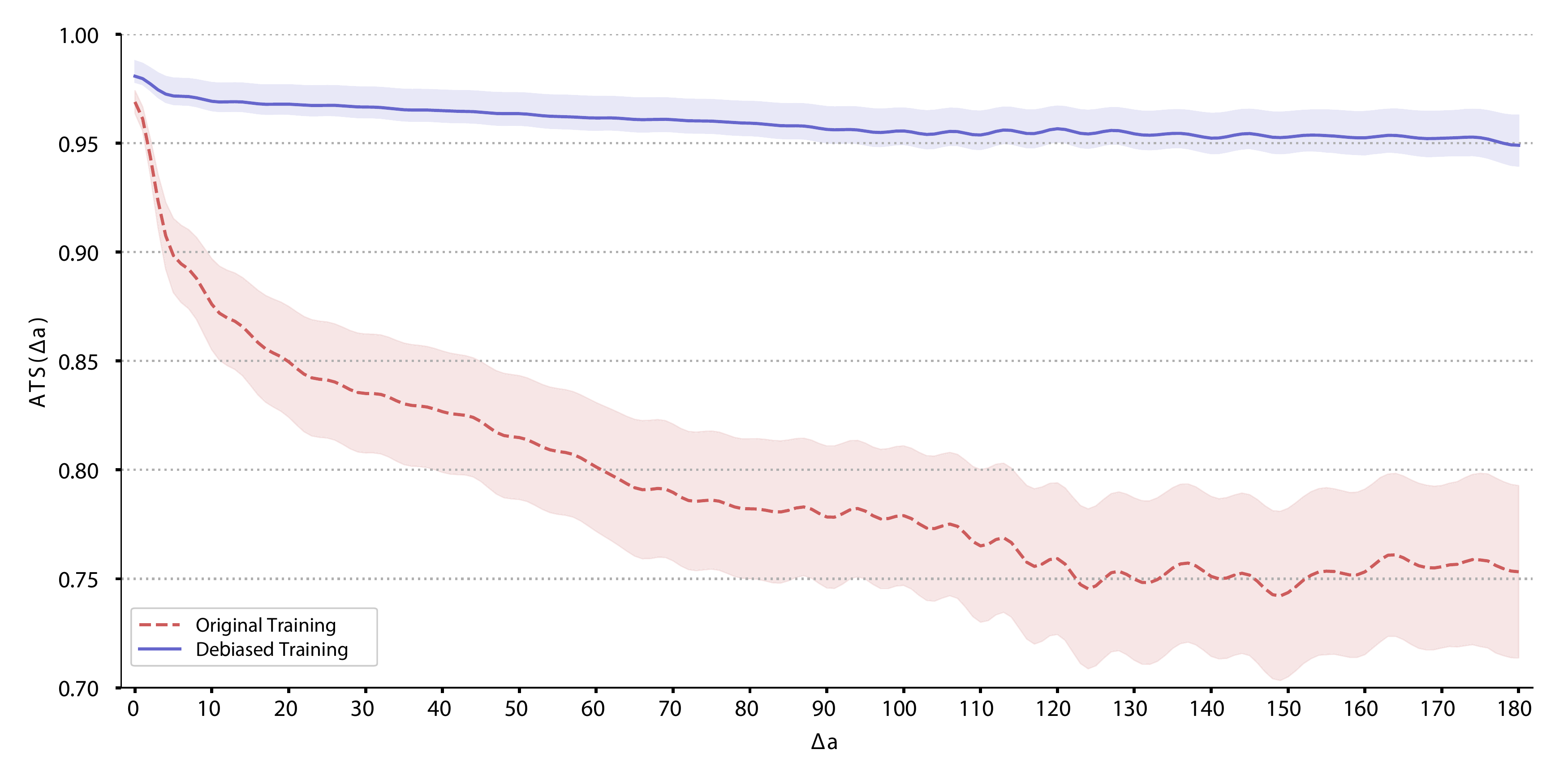}
    \centering
    \caption{ATS and standard deviation (y-axis) of same-term occurrences in different passages along positional $\Delta a$ of each term pair (x-axis) trained and evaluated on the MS MARCO collection with TK passage term representations.}
    \label{fig:position_bias_plot}
\end{figure*}

In Figure \ref{fig:position_bias_plot} we show the $\simf$ for different $(\Delta a)$ along the x-axis using TK passage term representations on the MS MARCO collection. The shaded area corresponds to the standard deviation. In this plot, an unbiased contextualization would result in a horizontal line, with a uniformly distributed standard deviation of the vectors. A set of contextualized vectors naturally has a standard deviation, as each vector, even for the same term is influenced by different context terms. 

It is evident from observing Figure \ref{fig:position_bias_plot} and Table \ref{tab:position_bias_stats}, that the TK model incurs a strong positional bias, especially for deltas smaller than 20. This shows the influence of the bias in the training data, which conditions the contextualized vectors on their absolute position. Using a debiased training set improves the representations and makes them much less dependent on their position. The SQuAD collection, not pictured in Figure \ref{fig:position_bias_plot}, exhibits a similar pattern, although dampened as the collection is less biased.

\section{Retrieval Results}
\label{sec:retrieval_results}

In this section we discuss our effectiveness related research questions with an emphasis on the differences in using the original vs. debiased training and evaluation, including the conclusion we can draw from them:

\vspace{0.2cm}
\noindent
\textit{\textbf{RQ2} What effect has the debiasing on the evaluation of Transformers?}
\vspace{0.1cm}

We look at the two collections separately to answer this RQ. In Table~\ref{tab:ir_biased_performance_marco} we have the results for the heavily-biased MS MARCO collection. We compare each measure by all possible training and evaluation approaches for all three Transformer models. The delta shows the relative difference between the original and debiased evaluation per training type. We can see that across all Transformer models we have a substantial drop in effectiveness when trained on the original training set and evaluated on the debiased set. This shows how the models learn to prioritize the beginning of the passages, and cannot generalize well to the scenario where answers are located in evenly distributed across the passage. The SQuAD results in Table~\ref{tab:ir_biased_performance_squad} on the other hand offer a different picture with only minor differences between original and debiased evaluation sets. This is to be expected, as we showed in Section \ref{sec:dataset_analysis} that the SQuAD collection is almost unbiased in its original form.

\begin{table*}[t!]
    \centering
    \caption{MSMARCO re-ranking results of original and debiased training sets (rows) on the original and debiased test sets (columns). Each measure uses a cutoff of 10 and the smallest absolute margin per block is marked in bold.}
    \label{tab:ir_biased_performance_marco}
    \setlength\tabcolsep{2.7pt}
    {
    \begin{tabular}{lr!{\color{lightgray}\vrule}ccr!{\color{lightgray}\vrule}ccr!{\color{lightgray}\vrule}ccr}
       \toprule
       \multicolumn{2}{c!{\color{lightgray}\vrule}}{\multirow{2}{*}{\textbf{Model}}}&
       \multicolumn{9}{c}{\textbf{MSMARCO - Test}} \\ 
       
       && \multicolumn{3}{c!{\color{lightgray}\vrule}}{\textbf{nDCG}}&
       \multicolumn{3}{c!{\color{lightgray}\vrule}}{\textbf{MRR}}&
       \multicolumn{3}{c}{\textbf{Recall}} \\
       &Training  & Orig. & Deb. & $\Delta$ & Orig. & Deb. & $\Delta$ & Orig. & Deb. & $\Delta$   \\
       \midrule
       
        \multirow{2}{*}{\textbf{BERT$_\textbf{CAT}$}} & Original  & 0.432 & 0.395 & -9.4\% & 0.372 & 0.336 & -10.7\% & 0.630  & 0.594 & -6.1\% \\
                                                      & Debiased & 0.416 & 0.415 & \textbf{-0.2\%} & 0.357 & 0.355 & \textbf{-0.6\%}  & 0.617  & 0.617 & \textbf{0.0\%} \\
         \arrayrulecolor{lightgray}
        \midrule
       \multirow{2}{*}{\textbf{BERT$_\textbf{DOT}$}} & Original  & 0.373 & 0.329 & -13.4\% & 0.316 & 0.276 & -14.5\% & 0.567 & 0.509 & -11.4\%  \\
                                                     & Debiased & 0.362 & 0.364 & \textbf{+0.6\%}  & 0.305 & 0.307 & \textbf{+0.7\%}  & 0.555 & 0.554 & \textbf{-0.2\%}  \\
        \midrule
       \multirow{2}{*}{\textbf{TK}} & Original  & 0.371 & 0.307  & -20.8\%  & 0.312 & 0.254 & -22.8\%   & 0.567 & 0.484 & -17.1\%  \\
                                    & Debiased & 0.356 & 0.355  & \textbf{-0.3\%}   & 0.298 & 0.296 & \textbf{-0.7\%}    & 0.551 & 0.552 & \textbf{+0.2\%}  \\

       \arrayrulecolor{black}
       \bottomrule
    \end{tabular}}
\end{table*}

\begin{table*}[t!]
    \centering
    \caption{Retrieval effectiveness results of original and debiased SQuAD training sets (rows) on the original and debiased SQuAD test sets (columns). Each measure uses a cutoff of 10 and the smallest absolute margin per block is marked in bold.}
    \label{tab:ir_biased_performance_squad}
    \setlength\tabcolsep{3pt}
    {
    \begin{tabular}{lr!{\color{lightgray}\vrule}ccr!{\color{lightgray}\vrule}ccr!{\color{lightgray}\vrule}ccr}
       \toprule
       \multicolumn{2}{c!{\color{lightgray}\vrule}}{\multirow{2}{*}{\textbf{Model}}}&
       \multicolumn{9}{c}{\textbf{SQuAD - Test}} \\ 
       && \multicolumn{3}{c!{\color{lightgray}\vrule}}{\textbf{nDCG}}&
       \multicolumn{3}{c!{\color{lightgray}\vrule}}{\textbf{MRR}}&
       \multicolumn{3}{c}{\textbf{Recall}} \\
       &Training  & Orig. & Deb. & $\Delta$ & Orig. & Deb. & $\Delta$ & Orig. & Deb. & $\Delta$   \\
       \midrule

        \multirow{2}{*}{\textbf{BERT$_\textbf{CAT}$}} & Original  & 0.908 & 0.902 & -0.7\% & 0.892   & 0.884 & \textbf{-0.9\%} & 0.957 & 0.956 & \textbf{-0.1\%} \\
                                                      & Debiased & 0.910 & 0.905 & \textbf{-0.6\%} & 0.894   & 0.885 & -1.0\% & 0.959 & 0.956 & -0.3\% \\
         \arrayrulecolor{lightgray}
        \midrule
       \multirow{2}{*}{\textbf{BERT$_\textbf{DOT}$}} & Original  & 0.780 & 0.783 & +0.4\% & 0.734   & 0.738 & +0.5\% & 0.924 & 0.919 & -0.5\% \\
                                                     & Debiased & 0.784 & 0.783 & \textbf{-0.1\%} & 0.740   & 0.739 & \textbf{-0.1\%} & 0.919 & 0.919 & \textbf{0.0\%} \\
        \midrule
       \multirow{2}{*}{\textbf{TK}} & Original  & 0.846 & 0.840 & -0.7\% & 0.818 & 0.811 & -0.9\% & 0.933 & 0.930 & -0.3\%  \\
                                    & Debiased & 0.848 & 0.844 & \textbf{-0.5\%} & 0.820 & 0.816 & \textbf{-0.5\%} & 0.932 & 0.931 &\textbf{ -0.1\%}  \\

       \arrayrulecolor{black}
       \bottomrule
    \end{tabular}}
\end{table*}

\vspace{0.2cm}
\noindent
\textit{\textbf{RQ3} Does a debiased training result in better generalization?}
\vspace{0.1cm}

In contrast to the poor original training to debiased test set results on MSMARCO in Table~\ref{tab:ir_biased_performance_marco}, using the debiased training set we observe similar results on the two test sets with little delta across all three models. These debiased training results are better than those using original training to debiased test sets, leading us to the conclusion that these results represent the true generalized effectiveness of the models. For the SQuAD results in Table~\ref{tab:ir_biased_performance_squad} we make an interesting observation, that some of the debiased trained models outperform those trained on the original training sets when applied to the original test sets. 

\vspace{0.2cm}
\noindent
\textit{\textbf{RQ4} Do we observe differences in transfer-learning, based on debiased pre-training?}
\vspace{0.1cm}

Finally, we look at a common transfer learning scenario: We utilize the large-scale MSMARCO as first retrieval pre-training and then transfer the trained model to a smaller collection (SQuAD) and train it again. This is especially helpful in production scenarios that require efficient models and do not provide ample training data. 

In Table \ref{tab:tl_results} we show our transfer learning results. We recall that the original MS MARCO is heavily biased and SQuAD is not. The debiased MS MARCO is closer to the SQuAD answer distribution. In general, using the MS MARCO pre-training improves the SQuAD results. For the production scenario models, that enable query independent passage representation caching -- BERT$_\text{DOT}$ and TK -- we observe another significant increase in effectiveness on SQuAD using the debiased MS MARCO training. Only BERT$_\text{CAT}$ does not benefit from the debiased pre-training.

\begin{table}[t!]
    \centering
    \caption{MS MARCO to SQuAD transfer learning results. Each measure uses a cutoff of 10. Significance is tested between training variants per model with Wilcoxon ($p<0.05$).}
    \label{tab:tl_results}
    \vspace{-0.2cm}
    \setlength\tabcolsep{6pt}
    {
    \begin{tabular}{lrc!{\color{gray}\vrule}lll}
       \toprule
       \multicolumn{3}{c!{\color{gray}\vrule}}{\multirow{1}{*}{\textbf{Model}}}&
       \multicolumn{3}{c}{\textbf{SQuAD Original Test}} \\
       & Train & Sig & nDCG & MRR & Recall \\
       \midrule
       
       \multirow{3}{*}{\textbf{BERT$_\textbf{CAT}$}} & SQuAD(Original)& $a$ & 0.908 & 0.892 & 0.957  \\  
       & MS(Original) $\rightarrow$ SQuAD(Original) & $b$  & \textbf{0.913} & \textbf{0.898} & 0.957  \\  
       & MS(Debiased) $\rightarrow$ SQuAD(Original) & $c$ & 0.911 & 0.896 & \textbf{0.958} \\   
       \arrayrulecolor{lightgray}
        \midrule
        \multirow{3}{*}{\textbf{BERT$_\textbf{DOT}$}} & SQuAD(Original)& $a$ & 0.780 & 0.734 & 0.924  \\  
       & MS(Original) $\rightarrow$ SQuAD(Original)   & $b$ & 0.788$^{a}$ & 0.744$^{a}$ & 0.922  \\  
       & MS(Debiased) $\rightarrow$ SQuAD(Original)  & $c$ & \textbf{0.792}$^{ab}$ & \textbf{0.748}$^{ab}$ & \textbf{0.927}$^{b}$ \\   
        \midrule
        
       \multirow{3}{*}{\textbf{TK}} & SQuAD(Original)& $a$ & 0.846 & 0.818 & 0.933  \\  
       & MS(Original) $\rightarrow$ SQuAD(Original)  & $b$ & 0.854$^{a}$ & 0.827$^{a}$ & 0.936  \\  
       & MS(Debiased) $\rightarrow$ SQuAD(Original) & $c$ & \textbf{0.857}$^{ab}$ & \textbf{0.832}$^{ab}$ & \textbf{0.937} \\   
       \arrayrulecolor{black}
        \bottomrule
    \end{tabular}}
    \vspace{-0.5cm}
\end{table}

\section{Related Work}
\label{sec:related_work}

\paragraph{Biases in datasets.}

Recent studies have observed a variety of artefacts (biases) in datasets of several NLP tasks. 
Gururangan et al. \cite{gururangan-etal-2018-annotation} demonstrate that for Natural Language Inference (NLI) datasets it is possible to identify the correct label by only looking at the hypothesis, without observing the premise based on superficial patterns generated while constructing the dataset. This is also confirmed by Poliak et al. \cite{poliak-etal-2018-hypothesis} and Tsuchiya et al. \cite{tsuchiya-2018-performance}. McCoy et al. \cite{mccoy-etal-2019-right} shows that state-of-the-art models follow simple heuristics to identify the correct answer.
Glockner et al. \cite{glockner-etal-2018-breaking} show the deficiency of state-of-the-art NLI architecture by testing them in an unbiased dataset.
Also QA and Visual QA (VQA) suffer from dataset artefacts. In fact, Jia and Liang \cite{jia-liang-2017-adversarial} show that human-level performance on SQuAD can be achieved by only relying on superficial cues, and Chen et al. \cite{chen-etal-2016-thorough} show that in NewsQA, 73\% of the answers can be predicted by simply identifying the single most relevant sentence. Formal et al. \cite{formal2020colbert} studied the reliance of the ColBERT \cite{khattab2020colbert} model on exact term matches in IR.

Another form of bias affecting IR test collections is the pool bias \cite{lipani2016,10.1007/978-3-319-30671-1_20}. This bias is a side effect of the sampling method used to build these test collections called, the pooling method \cite{8868196}. This is caused by the presence of non-annotated relevant documents in the collection which makes the evaluation of newly developed retrieval systems less reliable \cite{lipani2015,10.1145/2983323.2983891}.

Social biases are another form of bias manifesting in NLP and IR datasets \cite{doshivelez2017rigorous,gizem2021,rekabsaz2020neural}. In this case these biases are not generated by the way the datasets were constructed but by historical and cultural discriminations manifesting as a prejudice or unfair characterization of the members of a particular group. 

\paragraph{Bias mitigation methods.}

The research on the mitigation of these biases has branched out into two directions. One defining methods to mitigate biases when constructing the datasets. The other devising mechanism to make models robust against the presence of bias in datasets.  
Agrawal et al. \cite{Agrawal_2018_CVPR}, Anand et al. \cite{an2018blindfold}, and  Min et al. \cite{min-etal-2019-compositional} develop methods to build unbiased datasets without a variety of biases. 
Other forms of bias removal consist in learning unbiased representations. 
Bolukbasi et al. \cite{NIPS2016_6228} learned unbiased word embeddings to mitigate gender bias. 
Belinkov et al. \cite{belinkov-etal-2019-dont} propose two probabilistic methods to build models that are more robust to biases and better transfer across datasets.
Other methods to develop more robust NLP methods have been developed using adversarial methods \cite{clark-etal-2019-dont,li-etal-2018-towards,elazar-goldberg-2018-adversarial,barrett-etal-2019-adversarial,NIPS2018_7427,grand-belinkov-2019-adversarial}.
In the IR setting Gerritse et al. \cite{gerritse2020bias} studied and proposed methods to mitigate echo-chamber biases in personalised search. 

\paragraph{Modeling relative position in Transformers.}

To overcome this limitation in machine translation tasks, Shaw et al. \cite{shaw-etal-2018-self} developed a Transformer with a relation-aware self-attention, which induces the model to learn a relative positional encoding in a translation task. 
However, we have tested this Transformer-version and observed no improvement over the original version used in this paper. 
Also in translation tasks, Wang et al. \cite{wang-etal-2019-self} extend the transformer developed by Shaw et al. \cite{shaw-etal-2018-self} to model hierarchies based on a dependency tree.
We believe that these transformer-versions would benefit from our work, however we leave this to future work.

\section{Conclusion}
\label{sec:conclusion}

We observed a judgment bias towards the beginning of passages of selected answers in two popular QA datasets used for retrieval. Furthermore, the biased evaluation data hides the existence of this bias in the data. To overcome this problem, we proposed a dataset debiasing method, by switching two parts of a passage split at a random point, as the relevance of word matches in passage retrieval should be position independent. 

We showed how the excessive focus on earlier positions in the data propagates through Transformer-based contextualization to form position-biased representations. Our results show that three different Transformer ranking models (BERT$_\text{DOT}$, BERT$_\text{CAT}$, and TK) trained on the original (biased) MS MARCO collection, substantially lose effectiveness on the debiased version. On the SQuAD collection, acting as an unbiased control dataset, the models do not show this behavior.

We demonstrate that by using a debiased training data transformation, the Transformer models achieve the same performance on biased and debiased datasets, showing the increased generalizability of the models. Finally, we also show that for production-scenario transfer-learning, the debiased pre-training is the most effective strategy. This leads us to the conclusion that going forward, the community should adopt the simple data-transformation for debiasing the MSMARCO pre-training in these transfer-learning scenarios.

\bibliographystyle{splncs04}
\bibliography{main}

\end{document}